# On the Sieder state correction and its equivalent in mass transfer

Trinh, Khanh Tuoc

*K.T.Trinh@massey.ac.nz*

## Abstract

The physical background behind the success of the Sieder-Tate correction in heat transfer is analysed. The equivalent correction for mass transfer correlations is based on the ratio of diffusivities at the wall and bulk concentrations. This correction is not required if the Prandtl/Schmidt numbers are evaluated at the wall layer conditions and the Reynolds number at the bulk conditions. This technique brings heat and mass transfer coefficients into agreement.

## 1   Introduction

It is well-known that the heat transfer coefficients for cooling are somewhat lower than those for heating at the same Reynolds number. Sieder and Tate (1936) showed that they could be made to agree by introducing a correction factor into correlations of the form

$$Nu = A \operatorname{Re}^a \operatorname{Pr}^b \qquad (1)$$

The Sieder Tate correction factor $(\mu_w/\mu_b)^c$ is based on the observation that the Reynolds number for pipe flow for example

$$\operatorname{Re} = \frac{DV\rho}{\mu} \qquad (2)$$

where D is the pipe diameter, V the average fluid velocity and $\rho$ the density, is usually calculated with the viscosity $\mu_b$ measured at the temperature in the bulk flow but the viscosity at the wall $\mu_w$ is different during cooling and heating. The power index $c$ is often given the value *0.14*.

It has also been observed that mass transfer coefficients are some 10% different from heat transfer coefficients measured at the same Reynolds number (Harriott and Hamilton, 1965) as shown in Figure 1.

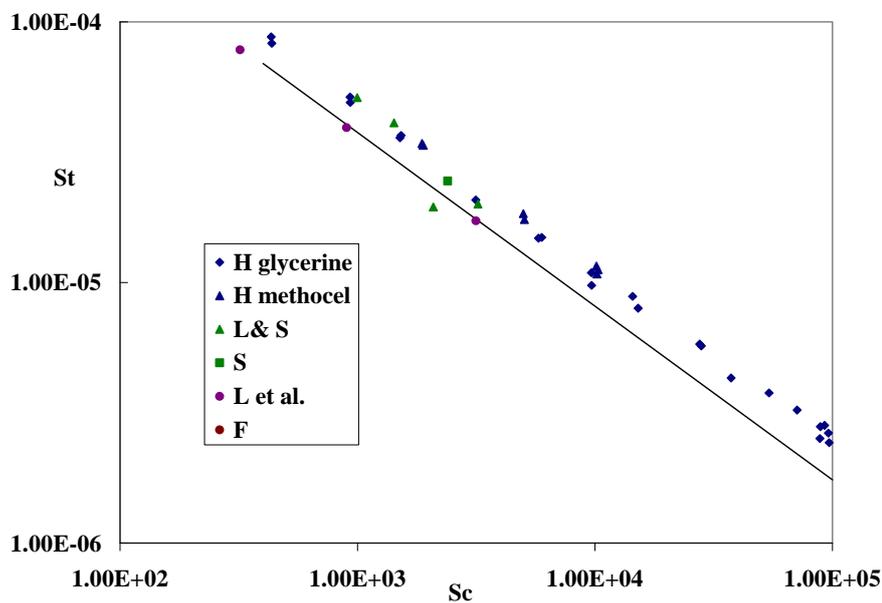

Figure 1. Stanton numbers for heat and mass transfer at Re $\approx$ *10,000*. Data H (Harriott and Hamilton, 1965), L&S (Linton and Sherwood, 1950), S (Shaw et al., 1963), L (Lin et al., 1953), F (Friend, 1958) Line predicted by (Metzner and Friend, 1958).

The present discussion presents a correction technique that I developed some time ago (Trinh, 1969) to deal with this discrepancy.

## 2    Theory

Many authors have also noted that most of the resistance to heat transfer occurs in a relatively thin wall layer where the effect of thermal diffusion are important. Deissler (1955) for example estimated the effect of temperature driven viscosity variations on the profile of velocity and temperature in the wall layer. This observation can be related to the wall layer sequence of inrush, sweep and ejection first observed by Kline et al. (1967). As argued in previous postings ((Trinh, 2009, 2010) the ejections bring lumps of all fluid from the wall that contain heat, mass and momentum into the outer region and form the physical basis of analogies between heat, mass and momentum transfer. This phenomenon results in the convection of a source of heat, mass and momentum that then diffuse into the surrounding fluid but the diffusivities of heat, mass and momentum do not contribute to the convection movement. They contribute mainly to the transfer from the wall to the adjacent fluid before it is ejected.

Penetration theories e.g. (Hanratty, 1956, Danckwerts, 1951, Harriott, 1962, Ruckenstein, 1968, Hughmark, 1968, McLeod and Ponton, 1977, Thomas and Fan, 1971, Loughlin et al., 1985, Fortuin et al., 1992, Hamersma and Fortuin, 2003, Kawase and Ulbrecht, 1983) further propose that the wall layer can be modeled in terms of an unsteady diffusion process and its scale can be calculated formally (Trinh, 2010). More classical boundary layer theories model the wall layer as steady or pseudo-steady state diffuse layers e.g. (Prandtl, 1910, Karman, 1930, Martinelli, 1947, Metzner and Friend, 1958, Reichardt, 1957, Levich, 1962, Churchill, 1977, 1997, Trinh, 1969)

All analogies begin with the application or Reynolds' analogy (1874) which can be written as

$$\frac{d\theta^+}{dy^+} = \frac{dU^+}{dy^+} \tag{3}$$

where the velocity $U^+ = U/u_*$, the normal distance from the wall $y^+ = yu_*\rho/\mu = yu_*/\nu$ and the temperature $\theta^+ = \rho C_p (\theta - \theta_w) u_*/q_w$ have been normalised with the friction velocity $u_* = \sqrt{\tau_w/\rho}$ and the fluid apparent viscosity $\mu$.

Equation (3) holds exactly for the region outside the wall layer (Trinh, 2010). If it is described in terms of a log-law first proposed by Prandtl (1935) and Millikan (1938) then the temperature profile can be expressed as

$$\Theta^+ = 2.5 \ln y^+ + B \tag{4}$$

The constant B is determined by forcing equation (4) through a point that defines the limit a thermal diffusion from the wall. The exact location of this boundary is the main point of difference between different analogies. Clearly it will depend on the value of the Prandtl number which must therefore be calculated at the conditions of the wall.

At the pipe axis, $y^+ = R^+$, $\Theta^+ = \Theta_m^+$ and $R^+ = (\text{Re}/2)\sqrt{f/2}$ and the Nusselt number can be derived from the temperature profile by standard techniques (e.g. Schlichting, 1960). An example correlation is (Trinh, 2010)

$$St = \frac{\sqrt{f/2}}{2.5 \ln \text{Re} \sqrt{f} + 15.55 \Pr^{2/3} - 13.03 + (2.5/3) Ln \Pr - D(\Pr, \text{Re})} \tag{5}$$

where

$$f = \frac{2\tau_w}{\rho V^2} \text{ is called the friction factor} \tag{6}$$

$$\text{Re} = \frac{DV\rho}{\mu} \text{ the Reynolds number} \tag{7}$$

$$St = \frac{h}{\rho C_p V} \text{ the Stanton number for heat transfer} \tag{8}$$

$$\Pr = \frac{C_P \mu}{k} \text{ the Prandtl number} \tag{9}$$

$\tau_w$ the wall shear stress, $C_p$ the heat capacity, $h$ the heat transfer coefficient and

$$D(\Pr, \text{Re}) = \Theta_m^+ - \Theta_b^+ \tag{10}$$

the difference between the maximum and bulk normalised temperature $\Theta^+$.

Thus the Reynolds number is introduced through the term $R^+$ and must be calculated at the bulk conditions. When both the Prandtl and Reynolds numbers are calculated at the bulk temperature, as was the custom before Sieder and Tate, then a correction factor is required. In equation (8) the viscosity is the variable most affected by temperature changes, which explains the success of the Sieder-Tate correction.

Equation (5) can be applied to mass transfer by substituting the Schmidt number

$$Sc = \frac{\mu}{\rho \mathcal{D}} \tag{11}$$

for the Prandtl number and the Stanton number for mass

$$St = \frac{k}{V} \tag{12}$$

for the Stanton number for heat with $k$ as the mass transfer coefficient and $\mathcal{D}$ the diffusivity of the transported material in the fluid.

In mass transfer studies, the driving force is the concentration difference across the boundary layer. Furthermore most studies use sparsely soluble material. Harriott and Hamilton (1965) stated that an "objective (of their study) was to vary the viscosity for some tests without changing the diffusivity as a check on the significance of the Schmidt number". The effect was not significant. Thus the Sieder-Tate correction cannot be applied in this situation.

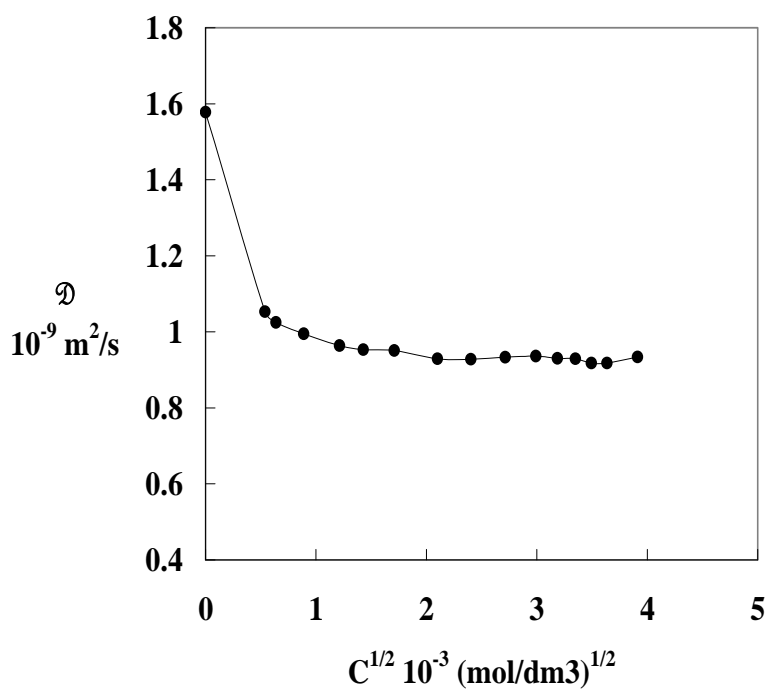

Figure 2. Variation of the diffusivity of benzoic acid in water with concentration. Data of Noulty and Lealst (1987).

In the Schmidt number, the diffusivity is the property most affected by concentration changes as shown in Figure 2. Since the diffusion wall layer is very thin compared to the pipe radius because of the high Prandtl/Schmidt number (Trinh, 1969, 2010), the Schmidt number must be evaluated at the average concentration in that layer in equation (5). When this is not done, a suitable correction factor is $(\mathcal{D}_w/\mathcal{D}_b)^d$ which is the mass transfer equivalent of the Sieder-Tate correction.

## 3 Comparison with experimental data and discussion

In the experiment of Harriott and Hamilton, the authors calculated the Schmidt number by taking "integral values of the diffusivity for a concentration range of about zero to almost saturation". Since the diffusion wall layer is very thin, this essentially means that the diffusivities relate to the bulk concentration. The authors report the solubility of benzoic acid as 0.0275 g-mol/l. My experimental results for analar benzoic acid was 3 g/l (Trinh 1969) which lies between the results of Harriott and Hamilton and Meyerink and Friedlander (1962).

For the range of Reynolds numbers between $10^4 - 10^5$, I estimated the average concentration in the wall layer to fall in the range $0.9 < C_f < 1.5 \, g/l$. In this range, the data of Chang (1949) and Read (1961) give a diffusivity value of $D_f = 0.86 \times 10^{-5} \, cm^2/s$ compared to the value of $0.961 \times 10^{-5} \, cm^2/s$ used by Harriott and Hamilton giving a ratio

$$\frac{Sc_f}{Sc_b} = \frac{\mathcal{D}_b}{\mathcal{D}_f} = 1.13$$

Application of an earlier correlation similar to equation (3) using Prandtl and Schmidt numbers estimated at average film temperature/concentrations (Trinh, 1969) gave the result in Figure 3. Several observations can be made from this figure

1. The use of the film Prandtl/Schmidt number instead of the bulk values brings heat and mass transfer coefficients in line.
2. Use of correlations such as equation (1) and (3) with $\Pr_f, Sc_f$ does not require any further correction factor $(\mu_w/\mu_b)^c$ for heat transfer, $(\mathcal{D}_w/\mathcal{D}_b)^d$ for mass transfer. They are only required when both the Reynolds and Prandtl/Schmidt numbers are calculated at the bulk conditions.

3. Equation (3) and similar formulations correlate experimental data adequately. We leave the full discussion of the different approaches to prediction of the heat and mass transfer coefficients and their levels of accuracy until further postings have been made to show how they can be applied to particular geometries and fluid rheological properties.

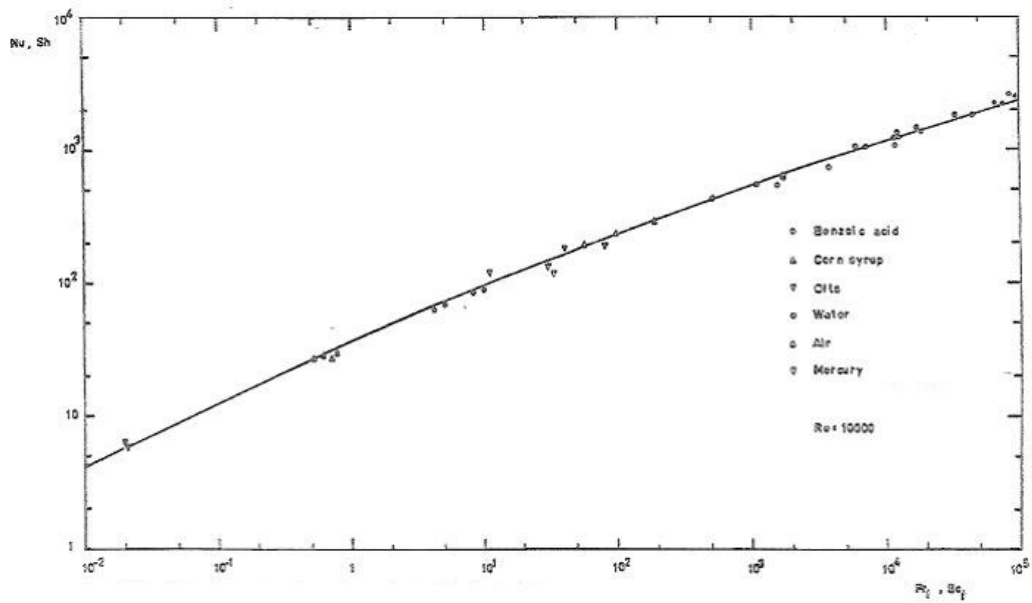

Figure 3. Variation of Nusselt and Sherwood numbers for heat and mass transfer at $Re = 10000$. Data: mass transfer - benzoic acid: Harriott and Hamilton (1962), Heat transfer: oil, corn syrup, water, Morrison and Whitman (1928); mercury, Isakoff and Drew (1951), Buhr et al (1968). Line represents an earlier version of equation 3 (Trinh, 1969).

## 4 Conclusion

The Prandtl number in heat transfer correlations must be evaluated at the average temperature in the wall layer and the Reynolds number at the bulk temperature. If both are evaluated at the bulk temperature, a Sieder-Tate type correction must be incorporated. For mass transfer the correction factor must be based on the ratio of diffusivities at the wall and bulk concentrations.